\documentclass[12pt]{article} 
\input{psfig}
\begin{document}

\author{J.C. Anjos\thanks{e-mail: janjos@cbpf.br}, J. Magnin
\thanks{e-mail: jmagnin@cbpf.br}\\
{\small Centro Brasileiro de Pesquisas F\'{\i}sicas} \\ 
{\small Rua Dr. Xavier Sigaud 150, CEP 22290-180, Rio de Janeiro, Brazil}\\
\\
G. Herrera
\thanks{e-mail: gherrera@fis.cinvestav.mx} \\ 
{\small Centro de Investigaci\'{o}n y de Estudios Avanzados} \\ 
{\small Apdo. Postal 14-740, M\'{e}xico 07000, DF, Mexico}
}

\title{On the intrinsic charm and the recombination mechanisms in charm hadron production}

\date{}
\maketitle

\begin{abstract}
We study $\Lambda_c^\pm$ production in $pN$ and $\pi^-N$ 
interactions. Recent experimental data from the
SELEX and E791 Collaborations at FNAL provide important 
information on the production mechanism of charm hadrons. 
In particular, the production of the $\Lambda_c$ baryon provides a 
good test of the intrinsic charm and the recombination 
mechanisms, which have been proposed to explain
the so called leading particle effects.
\end{abstract}

\newpage

\section{Introduction}

Hadronization of heavy quarks produced in hadron-hadron
interactions is still an open problem. The hadronization 
of quarks is in the realm of non-perturbative QCD and
not calculable from first principles yet. This is by far the less 
known aspect of heavy hadron production. 

The leading particle effect, which has been observed by several 
experiments in charm meson and baryon hadroproduction, indicates that 
charm hadronization cannot proceed by uncorrelated fragmentation 
alone. Furthermore, this effect implies the existence of some sort of 
recombination mechanism in the hadronization process.
Several models have been proposed to explain the leading particle effect, among them, the intrinsic 
charm mechanism~\cite{brodsky} and the conventional recombination two component 
model~\cite{nos}.

However, until now, no clear distinction has been made between the 
intrinsic charm and the conventional recombination mechanism. 
The reason is that there is no
experimental measurements of hadron antihadron production asymmetries 
and differential cross  sections simultaneously. 
As an exemption we can quote the WA82~\cite{wa82} 
and the WA92~\cite{wa92} experiments. These two experiments 
measured the $D^\pm$ and $D^\circ/\bar{D}^\circ$ asymmetries and 
differential cross sections as a function of $x_F$ 
($=2p_{||}/\sqrt{s}$). However, although there is some indications that 
the intrinsic charm model cannot describe simultaneously asymmetries 
and differential cross sections, these data are not conclusive as for the
production mechanism.

Recently, the SELEX Collaboration presented results on the $\Lambda_c^+-
\Lambda_c^-$ production asymmetries and particle distributions as a function of $x_F$ 
in the $p, \pi^-, \Sigma^--N\rightarrow \Lambda_c^\pm + X$ 
reactions~\cite{selex}. SELEX is a fixed target experiment with a beam average 
momentum of $600$ GeV/$c$. As we will show in the following, these results 
clearly favor the recombination over
the intrinsic  charm hypothesis as the dominant contribution to leading particle 
effects (see also Ref.~\cite{miguel}). 

Furthermore, E791~\cite{e791} results on $\Lambda_c^+-\Lambda_c^-$ production asymmetries
in $500$ GeV/$c$ $\pi^--N$ interactions seem 
to imply that other contributions like associated production of charm 
mesons should play a role in the observed asymmetry. In addition, it is 
interesting to note that the $\Lambda^\circ-\bar{\Lambda}^\circ$ 
asymmetry measured by the E791 Collaboration~\cite{e791b} is similar 
to the $\Lambda_c^+-\Lambda_c^-$ asymmetry, implying that the hadronization 
mechanisms for charm and strange quarks would be the same~\cite{prico}. 

In what follows, we will focus our attention in the intrinsic 
charm~\cite{brodsky} and recombination~\cite{nos} two component 
models as applied to the  $p-N\rightarrow \Lambda_c^\pm+X$ and
$\pi^--N\rightarrow \Lambda_c^\pm+X$ reactions.

\section{The differential cross section and asymmetries}

In two component models, the differential cross section is built on contributions 
from two different processes, namely, fragmentation of the 
heavy quarks (denoted {\it Frag}) and from the intrinsic charm
({\it IC}) or  recombination mechanisms ({\it Rec}),
\begin{equation}
\frac{d\sigma}{dx_F} = \frac{d\sigma^{Frag}}{dx_F} 
+ \frac{d\sigma^{IC(Rec)}}{dx_F} \; .
\label{eq1}
\end{equation}
The first term gives the same contribution to $\Lambda_c^+$ than to $\Lambda_c^-$ 
production, since no differences arise in charm or anticharm fragmentation. Contributions coming 
from the second term are, however, different for the $\Lambda_c^+$ and the $\Lambda_c^-$, thus generating 
a production asymmetry. The production asymmetry is defined as

\begin{equation}
A(x_F) =
\frac{d\sigma^{\Lambda_c^+}-d\sigma^{\Lambda_c^-}}
     {d\sigma^{\Lambda_c^+}+d\sigma^{\Lambda_c^-}}.
\end{equation}
\noindent
In a two components model, the production asymmetry would be given by
\begin{equation}
A(x_F) = \frac{d\sigma^{IC(Rec)}\left|_{\Lambda_c^+}-d\sigma^{IC(Rec)}\right|_{\Lambda_c^-}}
{2~d\sigma^{Frag}+d\sigma^{IC(Rec)}\left|_{\Lambda_c^+}+d\sigma^{IC(Rec)}\right|_{\Lambda_c^-}}\; .
\label{eq1b}
\end{equation}

In what follows we will analyse each contribution to eq~(\ref{eq1}), and hence to 
eq.~(\ref{eq1b}), separately.

\subsection{Fragmentation of heavy quarks}

The first component in eq.~(\ref{eq1}), which describes the production of heavy hadrons 
through the fragmentation of heavy quarks, is given by (see e.g. Ref.~\cite{nos})
\begin{equation}
\frac{d\sigma^{Frag.} }{dx_F}=\frac{1}{2} \sqrt{s} \int H^{AB}
(x_a,x_b,Q^2)
\frac{1}{E} \frac{D_{\Lambda_c/c} \left( z \right)}{z} dz dp_T^2 dy \; .
\label{eq2} 
\end{equation}
Here, $H^{AB}(x_a,x_b,Q^2)$ is given by
\begin{eqnarray}
H^{AB}(x_a,x_b,Q^2)& = & \Sigma_{a,b} \left[ q_a(x_a,Q^2)
\bar{q_b}(x_b,Q^2)\right. \nonumber \\
                   &   & + \left. \bar{q_a}(x_a,Q^2) q_b(x_b,Q^2) \right]
\frac{d \hat{\sigma}}{d \hat{t}} \left|_{q\bar{q}} \right. \nonumber \\
                   &   & + g_a(x_a,Q^2) g_b(x_b,Q^2) \frac{d \hat{\sigma}}
{d \hat{t}}\left|_{gg}\right. + ... 
\label{eq3}
\end{eqnarray}
and contain contributions from the pQCD processes $q\bar{q}\rightarrow c\bar{c}$, 
$gg\rightarrow c\bar{c}$, etc, and from the structure of the initial hadrons $A$ and $B$. 
$D_{\Lambda_c/c} ( z )$ is the Peterson fragmentation function given by~\cite{peterson}
\begin{equation}
D_{\Lambda_c/c}(z) = \frac{N}{z(1-1/z-\epsilon/(1-z)}\; ,
\label{eq1c}
\end{equation}
and $z$, $p_T^2$ and $y$ are the momentum fraction, the trasverse momentum and the rapidity of the 
heavy quark respectively. $x_a$ and $x_b$ are the momentum fractions of light quarks inside the 
initial hadrons $A$ and $B$.

As the fragmentation function~(\ref{eq1c}) is the same for $c$ and $\bar c$ fragmentation,
this term gives no  contribution to the production asymmetry of eq.~(\ref{eq1b}) at Leading
Order. At Next to Leading Order, 
a small $c-\bar c$ asymmetry translates into a tiny 
$\Lambda_c^+ - \Lambda_c^-$ asymmetry~\cite{nason}. 
However, this effect, which is very small, produces a negative asymmetry
in disagreement with experimental observations.

In our calculations we have used the GRV-LO parton distribution functions in proton and pions~\cite{grv}, 
$Q^2=4~m_c^2$, $m_c = 1.5$ GeV and $\epsilon=0.06$ in the Peterson fragmentation function.

\subsection{The intrinsic charm mechanism}

In $p-N\rightarrow \Lambda_c^\pm+X$ reactions, the intrinsic charm contribution 
comes from fluctuations of the beam protons to the 
$\left|uudc\bar{c}\right>$ Fock state, which breaks up in the collision 
contributing to $\Lambda_c^+$ production. 
The $\Lambda_c^+$ differential cross section for this process is~\cite{brodsky}
\begin{eqnarray} 
\frac{d\sigma^{IC} }{dx_F} & = & \beta 
\int_0^1dx_udx_{u^{\prime}}dx_ddx_cdx_{\bar{c} }\delta 
\left( x_F-x_u-x_d-x_c\right) \nonumber \\
                           &   & \times 
\frac{dP^{IC}}{dx_udx_{u'}...dx_{\bar c}} \; ,
\label{eq4} 
\end{eqnarray} 
where
\begin{equation}
\frac{dP^{IC}}{dx_udx_{u'}...dx_{\bar c}} = \alpha_s^4\left(M_{c
\bar{c}}^2\right) \frac{\delta \left(1-\Sigma _{i=u}^{\bar c} x_i
\right)}{\left( m_p^2 - \Sigma _{i=u}^{\bar c} \hat {m}_i^2 x_i \right)^2}
\: , 
\label{eq5} 
\end{equation}
is the probability of the $\left|uudc\bar{c}\right>$ fluctuation of the 
proton and $\beta$, which gives the probability of the fluctuation, is a parameter 
which must be fixed 
adequately to describe experimental data. 

To obtain a $\Lambda_c^-$ in $p-N$ interactions, a fluctuation of the proton to the 
$\left|uudu\bar{u}d\bar{d}c\bar{c}\right>$ Fock state is required. Since the probability of a 
five-quarks state is larger than for a nine-quarks Fock state, $\Lambda_c^+$ production is 
favored over $\Lambda_c^-$ in proton initiated reactions. A similar expression to those of
eqs.~(\ref{eq4}) and  (\ref{eq5}) can be found for the $\Lambda_c^-$ differential cross
section. However, its contribution is negligible.

A similar mechanism is at work in $\pi^--N$ interactions. However, $\Lambda_c^\pm$ production 
in $\pi^-$ initiated reactions requires fluctuations of the pion to
$\left|\bar{u}d\bar{u}u\bar{c}c\right>$ or  
$\left|\bar{u}d\bar{d}dc\bar{c}\right>$ Fock states. Then, after the break-up, a 
$\Lambda_c^++\Sigma_c^{--}$ or $\Lambda_c^-+\Sigma_c^0$ is formed, respectively, in the final state. 
Since the invariant mass of both final states should be approximately the same, 
the contribution to $\Lambda_c^+$ is the 
same than for $\Lambda_c^-$ production and no asymmetry at all is obtained.

\subsection{The recombination mechanism}

The conventional recombination contribution to the second term of eq.~(\ref{eq1}) in $p-N$ interactions  
has the form~\cite{nos}
\begin{equation}  
\frac{d\sigma^{Rec.}}{dx_F}= \beta
\int_0^{1}\frac{dx_1}{x_1}\frac{dx_2}{x_2}
\frac{dx_3}{x_3}F_3^p\left( x_1,x_2,x_3\right) 
R_3\left( x_1,x_2,x_3,x_F\right) \;,
\label{eq6} 
\end{equation} 
where $F_3^p\left( x_1,x_2,x_3\right)$ is the multiquark distribution and 
$R_3\left( x_1,x_2,x_3,x_F\right)$ the recombination function. As 
for the intrinsic charm model, $\beta$ is a parameter which must be fixed 
from experimental data. $x_i$ (i=1,2,3) are 
the momentum fractions of quarks in the initial proton which will be 
valence quarks in the final $\Lambda_c^+$ ($\Lambda_c^-$).

The recombination model assumes that there exist charm 
quarks inside the proton. The process of charm 
production occurs at a scale of the order of $Q^2\sim4m_c^2$, which is above 
the threshold for the perturbative production of charm inside the 
proton~\cite{halzen}. The charm inside the proton can have 
both, a non-perturbative and a perturbative origin due to QCD evolution, 
with the first existing over a scale independent of $Q^2$. However, 
for $Q^2\sim4m_c^2$  the perturbative component must be dominant.

Leading particle effects in the recombination model are due to 
the different contributions to the multiquark distribution. Actually, 
for $\Lambda_c^+$ production in $p-N$ interactions
\begin{equation}
F_3^p\left( x_1,x_2,x_3\right) =
x_1u^p(x_1)x_2d^p(x_2)x_3c^p(x_3)\rho(x_1,x_2,x_3)\; ,
\label{eq7}
\end{equation}
while for $\Lambda_c^-$ production
\begin{equation}
F_3^p\left( x_1,x_2,x_3\right) =
x_1\bar{u}^p(x_1)x_2\bar{d}^p(x_2)x_3\bar{c}^p(x_3)\rho(x_1,x_2,x_3)\; .
\label{eq8}
\end{equation}
The multiquark distribution given in eq.~(\ref{eq7}) recibes 
contributions from valence and sea quarks in the proton whereas the 
multiquark distribution in eq.~(\ref{eq8}) has contributions coming 
from the sea of the proton alone. $\rho(x_1,x_2,x_3)$ in eqs.~(\ref{eq7}) 
and (\ref{eq8}) correlates in momentum 
the single quark distributions. We used~\cite{nos}
\begin{equation}
\rho(x_1,x_2,x_3)=(1-x_1-x_2-x_3)^{-0.1}
\label{eq10}
\end{equation}
for both, $\Lambda_c^+$ and $\Lambda_c^-$ production.

For the recombination function we simply used~\cite{nos}
\begin{equation}
R_3\left( x_1,x_2,x_3\right) =\alpha \frac{x_1x_2x_3}
{x_F^2}\delta \left(x_1+x_2+x_3-x_F\right)\; ,
\label{eq9}
\end{equation}
with the parameter $\alpha$ in eq.~(\ref{eq9}) fixed by the condition~\cite{das-hwa}
\begin{equation}
\int_0^1{dx_1dx_2dx_3\frac{R(x_1,x_2,x_3,x_)}{x_F^3}}= 1\; .
\label{eq10b}
\end{equation}

In $\pi^--N\rightarrow \Lambda_c^\pm+X$, the differential cross section and the 
recombination function 
are given by expressions formally identical to those of eqs.~(\ref{eq6}) and (\ref{eq9}). 
However, the 
multiquark distribution function is different for $\Lambda_c^+$ and $\Lambda_c^-$ 
production. 
In fact, for $\Lambda_c^+$ we have 
\begin{equation}
F_3^{\pi^-}(x_1,x_2,x_3) = x_1d^\pi(x_1)x_2u^\pi(x_2)x_3c^\pi(x_3)\rho(x_1,x_2,x_3)\; ,
\label{eq101}
\end{equation}
while for $\Lambda_c^-$ it is
\begin{equation}
F_3^{\pi^-}(x_1,x_2,x_3) = 
r x_1\bar{d}^\pi(x_1)x_2\bar{u}^\pi(x_2)x_3\bar{c}^\pi(x_3)\rho(x_1,x_2,x_3)\; ,
\label{eq102}
\end{equation}
here $r$ is a suppression factor lower than one. For the $\rho$ function we used the same as
in eq.~(\ref{eq10}). 

The origin of the suppression factor $r$ in eq.~(\ref{eq102}) can be understood as follows: 
for $\Lambda_c^-$ production, the multiquark distribution is built up from the $\bar u$, $\bar d$ 
and $\bar c$ quark distributions in the pion. But, $\bar u$ and $\bar d$ quarks in the pion
can easily annihilate 
with $u$ and $d$ valence quarks in the nucleon, thus reducing the amount of $\bar u$ and $\bar d$ 
quaks in the pion available to recombine into a $\Lambda_c^-$. This suppression is not present 
in $\Lambda_c^+$ production since $u$ and $d$ quarks in the pion can only annihilate with
$\bar u$ and $\bar d$ 
{\em sea} quarks in the nucleon.

\subsection{Comparison to experimental data}

\begin{figure}[b] 
\psfig{figure=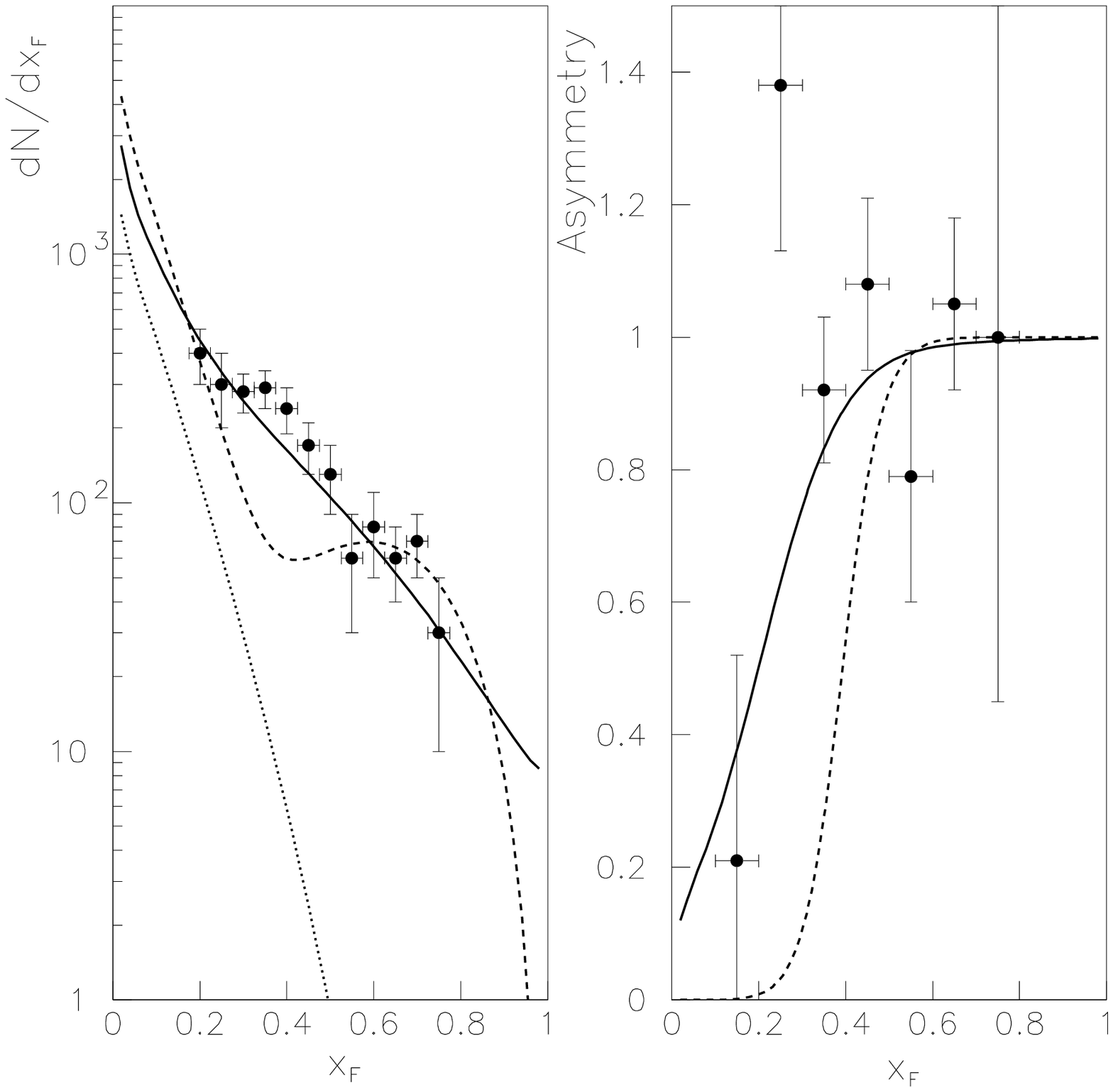,height=6.0in} 
\caption{Differential cross section (left) and production asymmetry (right) for $\Lambda_c^\pm$ 
production in $p-N$ interactions. Experimental data were taken from Ref.~\cite{selex}. 
The solid line shows the prediction of the {\em Rec} two component model. The {\em IC} prediction 
is shown by the dashed line. The dotted line shows also the contribution from Peterson fragmentation to the 
total cross section in the {\em Rec} model.}
\label{fig1} 
\end{figure}

\begin{figure}[b] 
\psfig{figure=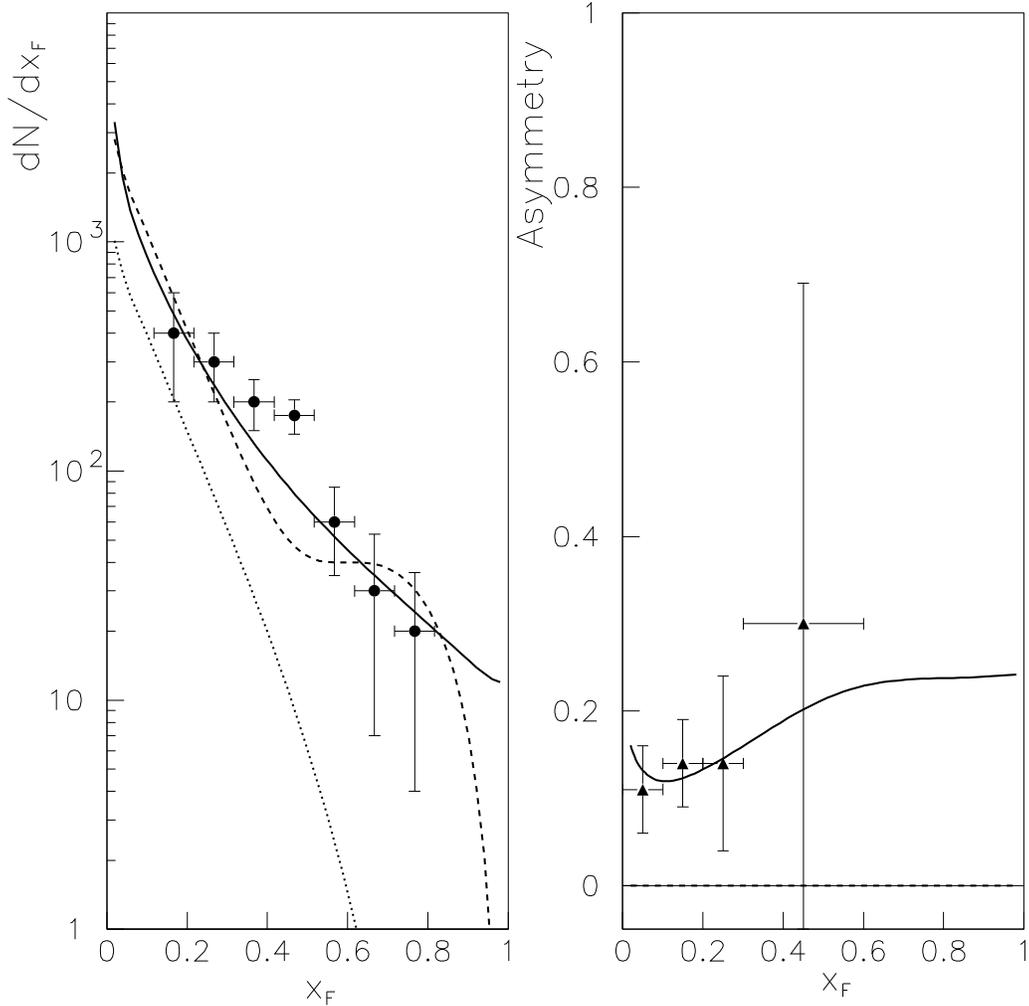,height=6.0in}
\caption{Differential cross section (left) and production asymmetry (right) for $\Lambda_c^\pm$ 
production in $\pi^--N$ interactions. Experimental data on the differential cross section were taken from 
Ref.~\cite{selex} while data on asymmetry are from Ref.~\cite{e791}. The SELEX experiment 
has also measured the $\Lambda_c$ production asymmetry in $\pi^--N$ interactions, but 
with error bars larger  than those of the E791 experiment. 
The solid line shows the prediction of the {\em Rec} two component model. The {\em IC} prediction 
is shown by the dashed line. The dotted line shows also the contribution from Peterson fragmentation to the 
total cross section in the {\em Rec} model.}
\label{fig2} 
\end{figure}

In Figs.~(\ref{fig1}) and (\ref{fig2}) we compare predictions of the {\em IC} and 
{\em Rec} models to experimental 
data from SELEX~\cite{selex} and E791~\cite{e791} experiments on 
$p-N\rightarrow \Lambda_c^{\pm}+X$ and 
$\pi^--N\rightarrow \Lambda_c^{\pm}+X$ respectively. 

In order to fix the parameters in both models, we used
\begin{equation}
\frac{d\sigma}{dx_F} = N \left[\frac{d\sigma^{Frag}}{dx_F}+\beta\frac{d\sigma^{IC(Rec)}}{dx_F}\right]\; ,
\label{last}
\end{equation}
where $N$ is a global normalization factor. The individual cross sections for each contribution have 
been normalized to unity, except for the $\Lambda_c^-$ distribution in recombination in $p-N$ 
interactions, 
which has been normalized by means of $N_{\Lambda_c^-}=1/\int{\frac{d\sigma_{\Lambda_c^+}}{dx_F}dx_F}$ to preserve 
the relative amount among $\Lambda_c^+$ and $\Lambda_c^-$ production. 

In $\pi^--N$ interactions the $\Lambda_c^+$ and $\Lambda_c^-$ recombination cross sections were 
normalized to unity. For the last one, the $r$ factor was included in the definition of the parameter $\beta$ 
in eq.~(\ref{last}).

In this way we can have an approximate idea of the relative size of each contribution to the total 
cross section.

Furthermore, in order to have the curves shown in the figures, the parameters in the {\em IC} and 
{\em Rec} models were fixed to values which best describe the differential cross sections. 
Once this was done, the asymmetry was calculated. The $r$ parameter was fixed in order to have a 
good description of the asymmetry in $\pi^--N$ interactions.

For $\Lambda_c^\pm$ production in $p-N$ interactions we used $\beta=1.8~(\beta=0.1)$ in the 
{\em Rec~(IC)} model , indicating that recombination is a substantial part of $\Lambda_c$ production. 
The same value for the $\beta$ parameter was used for the {\em Rec} model in $\pi^--N$ interactions, 
but a slightly lower $\beta=0.06$ was used for the {\em IC} model. A suppression factor $r=0.6$ is required to 
describe the production asymmetry in the {\em Rec} model. 

It must be noted that equally good descriptions for both, the differential cross section and asymmetry, can be 
obtained in the framework of the {\em Rec} model using values for $\beta$ in the range $1-2$ 
and varying the global normalization factor $N$ in eq.~(\ref{last}) accordingly. This means that parameters 
in the {\em Rec} model can only be fixed with accuracy once data for the differential cross section on 
$\Lambda_c$ production in the low $x_F$ region ($0<x_F<0.2$) become available.

\section{Conclusions}

For the first time, experimental data on $\Lambda_c$ production and production asymmetries
allow to  distinguish among two different mechanisms of production and hadronization.
It seems that the {\em IC} two components model do not describe simultaneously the
differential 
cross sections and production asymmetries for $\Lambda_c^\pm$ produced in $p-N$ and 
$\pi^--N$ interactions. 
Conversely, the {\em Rec} two component model seems to be a sensible approach to the problem, 
giving a good description of both, the $\Lambda_c$ differential cross section and the production 
asymmetry.

In addition, we have shown that the {\em Rec} two component model is able to explain the 
positive asymmetry 
observed by the E791~\cite{e791} and SELEX~\cite{selex} experiments in $\pi^--N$ interactions. 
As discused in  the text, the {\em IC} model predicts none asymmetry in this case.

Furthermore, the recombination mechanism seems to be more important than fragmentation. In fact, in 
$p-N\rightarrow\Lambda_c+X$ the recombination contributions is 1 to 2 times bigger than fragmentation. The same 
is observed in $\pi^--N\rightarrow\Lambda_c+X$. This is a clear signal that the debris of the initial hadrons 
play a fundamental role in the hadronization process.

In Ref.~\cite{gutierrez}, predictions in the framework of the {\em IC} model have been done on the 
$\Lambda_c^\pm$ production and asymmetry in $\Sigma^--N$ interactions. Although no comparison to experimental 
data is made a extremely hard behaviour is seen in the curves for the differential cross section.

\section*{Acknowledgements}

This work was supported by CONACyT, Mexico and CNPQ, Brazil. J.M. gratefully acknowledges 
the kind hospitality at CINVESTAV-Mexico, where part of this work was done.

\newpage

\newpage

\end{document}